\documentstyle[aps,oldlfont,amstex,amssymb,eqsecnum]{revtex}

\newtheorem{thm}{Theorem}
\newtheorem{lem}{Lemma} 
\newtheorem{rem}{Remark} 

\makeatletter
\def\dddot#1{{\mathop{#1}\limits^{\vbox to-1.4\ex@{\kern-\tw@\ex@
\hbox{\series m\normalshape...}\vss}}}}
\def\ddddot#1{{\mathop{#1}\limits^{\vbox to-1.4\ex@{\kern-\tw@\ex@
\hbox{\series m\normalshape....}\vss}}}}
\makeatother

\newcommand{\res}{\mathop{\rm res}\nolimits}
\newcommand{\const}{\mathop{\rm const.}\nolimits}
\newcommand{\pr}{\mathop{\rm pr}\nolimits}
\newcommand{\Sl}{\mathop{\rm sl}\nolimits}
\newcommand{\gl}{\mathop{\rm gl}\nolimits}
\newcommand{\lhs}{\mathop{\rm l.h.s.}\nolimits}
\newcommand{\rhs}{\mathop{\rm r.h.s.}\nolimits}

\newcommand{\Pol}{\mathop{\rm Pol}\nolimits}

\newenvironment{pf}{ \par\topsep6pt plus6pt
\trivlist \item[\hskip\labelsep\it
Proof.]\ignorespaces}{$\square$\endtrivlist}

\begin{document}
\title{Loop Algebra Symmetries and Commuting Flows for the
Kadomtsev-Petviashvili Hierarchy}
\author{A. Yu. Orlov\thanks{Permanent address: Institute of
Oceanology, Krasikova 23, Moscow 117218, Russia}}
\address{Centre de Recherches Math\'ematiques, Universit\'e de
Montr\'eal, Case postale~6128 -- succursale centre-ville, Montr\'eal
(Qu\'ebec) H3C 3J7, Canada}
\author{P. Winternitz}
\address{Centre de Recherches Math\'ematiques, Universit\'e de
Montr\'eal, Case postale~6128 -- succursale centre-ville, Montr\'eal
(Qu\'ebec) H3C 3J7, Canada}
\maketitle
\begin{abstract}
The relation between the $\widehat\gl(\infty)$ symmetry of the
Kadomtsev-Petviashvili hierarchy and the Kac-Moody-Virasoro Lie point
symmetries of the individual equations is established. The Lie point
symmetries are shown to be the only local ones.
\end{abstract}

\section{Introduction}

The purpose of this article is to perform an exhaustive study of all
symmetries of the Kadomtsev-Petviashvili (KP) hierarchy of partial
differential equations (PDE). We shall use the word ``symmetry'' in a
very general sense that will include point symmetries, higher
symmetries and also nonlocal ones. Indeed, a ``symmetry'' of an
equation, or a set of equations, will be any further equation,
compatible with the studied one. In the context of integrable
evolution equations we shall understand ``symmetries'' as flows,
commuting with the considered flow.

The Kadomtsev-Petviashvili equation~\cite{ref1} was introduced in the
context of waves propagating in shallow water, or in a plasma, and
corresponds to small transverse perturbations of solutions of the
Korteweg-deVries equation. It is currently under intense study in
view of its applications in conformal quantum field theory and string
theory~\cite{ref2}.

The KP equation is known to be integrable, in the sense that there is
a Lax pair associated with it, it allows infinitely many conservation
laws, solitons and has all the usual attributes of
integrability~\cite{ref3}--\cite{ref6}. The KP equation is the first
nontrivial member of an infinite hierarchy of mutually compatible
equations,~\cite{ref7,ref8} each representing a flow with respect to a
different ``times'' $t_n$.

Two different types of infinite dimensional Lie groups are associated
with this hierarchy. One is the group $\gl(\infty)$, figuring
prominently in the work of Jimbo and Miwa~\cite{ref7}. It acts on
$\tau$-functions, providing solutions to an entire infinite hierarchy
of equations. The corresponding group transformations are in general
nonlocal: they involve integrals of solutions.

A general approach to the study of symmetries of integrable
hierarchies of equations was developed by Orlov and
Schulman~\cite{ref9}, \dots ,\cite{ref12}. We also mention work by
Fuchssteiner~\cite{ref13} and Chen et al.~\cite{ref14} on symmetries
of the KP hierarchy.

The other infinite dimensional Lie group associated with the KP
equation is the local Lie group of local point transformations taking
solutions of the equation into solutions. It is this group that is
usually called the ``symmetry group'' of an equation and its Lie
algebra is the ``symmetry algebra''. The calculation of the symmetry
group of a differential equation is entirely algorithmic~\cite{ref15}
and can be done using various computer packages~\cite{ref16}.

It turns out that for the KP equation the corresponding symmetry
algebra has a characteristic Kac-Moody-Virasoro
structure~\cite{ref17,ref18}. Moreover, the Kac-Moody-Virasoro
symmetry algebras are typical for all integrable equations involving
three independent variables, such as the Davey-Stewartson equation,
the 3-wave resonant interaction equations, and many
others~\cite{ref19,ref20,ref21}. The fact that the symmetry groups of
the KP equation and the potential KP equation are infinite dimensional
was also observed by Schwarz~\cite{ref22} and by Infeld and
Frycz~\cite{ref23}.

The purpose of this article is to establish the relation between the
$\gl(\infty)$ symmetries of the KP hierarchy and the Lie point
symmetries of individual equations in the hierarchy, in particular the
KP equation itself. In the process we will establish several further
results. In particular we show that the Lie point symmetry algebra of
each equation in the KP hierarchy has a Kac-Moody-Virasoro structure.
A further result is that the Orlov-Schulman
symmetries~\cite{ref9},\dots ,\cite{ref12}, sometimes called
$W_{\infty}$ symmetries, correspond to Lie point symmetries, whenever
they are local. Finally, we clarify the meaning of the ``conditional
symmetries'' of ``conditionally integrable'' evolution equations,
considered earlier~\cite{ref24,ref25}.

In Section II we review some results on the KP hierarchy and its
symmetries, making use of the Gel'fand-Dickey approach via the algebra
of pseudodifferential operators~\cite{ref8}. We present the results
in a unified manner and include some known, but not easily accessible
results~\cite{ref9,ref10,ref11}.

Section III is devoted to the Lie point symmetries of the first two
equations in the KP hierarchy. The main results, namely the relation
between $W_{\infty}$ and Lie point symmetries are derived in Section
IV.

\section{The $KP$ hierarchy and its generalized symmetries}

\subsection{The algebra of pseudodifferential operators}

We shall make use of the algebra of pseudodifferential operators (PDO)
in one variable $f(x, \partial)$ satisfying the permutation rule
\begin{equation}
f(\partial) g(x) =
	\sum_{k=0}^\infty \frac{g^{(k)} (x) f^{(k)} (\partial)}{k!} ,
\label{eq2.1}
\end{equation}
where $(k)$ denotes the $k$-th derivative with respect to the
argument. For example, we have
$$
\partial^{-1} x =
	x \partial^{-1} - \partial^{-2}.
$$
For a detailed exposition see e.g. Ref. 8.

An operation of conjugation $(*)$ is introduced, defined by the rules
\begin{equation}
x^* = x, \quad \partial^* = - \partial, \quad (AB)^* = B^*A^*.
\label{eq2.2}
\end{equation}

We shall also make use of the splitting of the space of PDO into the
direct sum of two linear spaces, $A = A_+ \dotplus A_-$, with
\begin{equation}
\begin{aligned}
( \Sigma a_n \partial^n)_+	&= \sum_{n \ge 0} a_n \partial^n ,\\
( \Sigma a_n \partial^n)_- &= \sum_{n < 0} a_n \partial^n .
\end{aligned}
\label{eq2.3}
\end{equation}
We see that $A_+$ is the subalgebra of differential operators, $A_-$
the subalgebra of purely integral operators.

Let us introduce an integral operator $K$ with variable coefficients
\begin{equation}
K = 1 + \sum_{n=1}^\infty K_n (x) \partial^{-n}
\label{eq2.4}
\end{equation}
($K$ can be called the formal Zakharov-Shabat dressing operator, by
analogy with the analytical dressing operator of Ref. 4). If the
functions $K_n(x)$ are sufficiently smooth, an inverse integral
operator exists, namely
\begin{equation}
K^{-1} = 1 + \sum_{n=1}^\infty \widetilde{K}_n(x) \partial^{-n},
\label{eq2.5}
\end{equation}
with coefficients
\begin{equation}
\widetilde{K}_n(x) =	-K_n(x) + P(K_1, K_2, \dots , K_{n-1}),
\label{eq2.6}
\end{equation}
where $P$ is a differential polynomial in the indicated arguments.

We shall also need the $*$-conjugate operator
\begin{equation}
K^{*-1} = 1 + \sum_{n=1}^\infty K^*_n (x) \partial^{-n}
\label{eq2.7}
\end{equation}
(where $K^*_n(x)$ is simply a notation for new coefficients, not
$*$-conjugates of $K_n(x)$).

\subsection{Vector fields and their commutators}

Let us now consider the space of pesudodifferential operators of the form
\begin{equation}
C_i =	\sum_n c_n^i(x, t_1, t_2, t_3, \dots)\partial^n,
\label{eq2.8}
\end{equation}
where $c_n$ are some fixed functions. The ``time'' $t_i$ plays a
privileged role in the operator $C_i$ and in the coefficients $c_n^i$.
Note that the summation can be over both positive and negative values
of $n$. Let us now introduce a mapping
\begin{equation}
C_i \to V_{t_i}
\label{eq2.9}
\end{equation}
from the PDOs $C_i$ onto vector fields $V_{t_i}$ acting on the
integral operators $K$ of eq.~(\ref{eq2.4}) according to the rule
\begin{equation}
V_{t_i} K \equiv \biggl[ \frac{\partial}{\partial t_i}, K \biggr] =
-(KC_i K^{-1})_- K.
\label{eq2.10}
\end{equation}

Each vector field $V_{t_i}$ induces a flow of all of the coefficients
$K_n(x, \vec t)$ of $K$, once we allow the coefficients $K_n(x)$ in
eq.~(\ref{eq2.4}) to depend on $\vec t$, as well as on $x$.

A result that will be used a great deal below is the following.

\begin{thm} \label{thm1} The commutator of two vector fields of the
form of eq.~{\em(\ref{eq2.10})} satisfies
\begin{equation}
[V_{t_i}, V_{t_j}]K = -(KC_{ij} K^{-1})_- K,
\label{eq2.11}
\end{equation}
where we have
\begin{equation}
C_{ij} = [C_i, C_j] + \frac{\partial C_i}{\partial t_j} -
\frac{\partial C_j}{\partial t_i}.
\label{eq2.12}
\end{equation}
\end{thm}

\begin{pf} Let eq.~(\ref{eq2.10}) hold for $V_{t_i}$ and $V_{t_j}$.
We then have
$$
V_{t_i}(V_{t_j} K)	= V_{t_i} (-KC_j K^{-1})_- K.
$$
Since $V_{t_i}$ acts as a differentiation, and commutes with the
splitting (\ref{eq2.3}), we obtain
$$
V_{t_i} V_{t_j} K = -\bigl( (V_{t_i} K) C_j K^{-1} \bigr)_- K - \bigl(
K(V_{t_i} C_j) K^{-1} \bigr)_- K +
\bigl( KC_j K^{-1} (V_{t_i}K)K^{-1}\bigr)_- K - (KC_j K^{-1})_- V_{t_i} K.
$$
We replace $V_{t_i}K$ using eq.~(\ref{eq2.10}) again and then write
the commutator as
\begin{multline*}
V_{t_i}V_{t_j} K - V_{t_j}V_{t_i} K =	\bigl( K(V_{t_j} C_i -
V_{t_i}C_j) K^{-1} \bigr)_- K
	+ \Bigl\{ \bigl[ (KC_iK^{-1})_-,(Kc_jK^{-1}) \bigr] \\
- \bigl[ \bigl(KC_iK^{-1} \bigr)_-,KC_iK^{-1} \bigr] + \bigl[
(KC_jK^{-1})_-, (KC_i K^{-1})_- \bigr] \Bigr\}_- K.
\end{multline*}
The term in curly brackets can be simplified, using the decomposition
(\ref{eq2.3}) and the fact that we have $\bigl( ( \ \ )_+ ( \ \ )_+
\bigr)_- = 0$. We obtain
$$
\{ \ \ \}_-
	= \Bigl\{ \bigl[ (KC_i K^{-1}), (KC_j K^{-1} \bigr] \Bigr\}_-
	= \Bigl\{ K[C_i, C_j]K^{-1}\Bigr\}_-
$$
and this completes the proof.
\end{pf}

\begin{rem}\normalshape
\begin{enumerate}
\item[1.] The operators $C_i = C^i_{nm} (\tau) x {}^n\partial^m$ can
be viewed as connections corresponding to the algebra of PDO and
eq.~(\ref{eq2.12}) defines a curvature. Eq.~(\ref{eq2.12}) shows that
the mapping (\ref{eq2.10}) is not a mapping of the algebra of PDO, to
the algebra of vector fields $V$. The zero curvature condition
$c_{ij} = 0$ means that the flows with respect to $\tau_i$ and
$\tau_j$ are compatible.
\item[2.] A supersymmetric version of eq.~(\ref{eq2.11}) has been used
to construct supersymmetries for the super KP hierarchy~\cite{ref9}.
\end{enumerate}
\end{rem}

\subsection{The KP hierarchy}

Let us again consider the integral operator $K$ of eq.~(\ref{eq2.4})
and this time interpret the coefficients $K_n$ as depending on an
infinite sequence of times $t_1, t_2, t_3, \dots$. We shall identify
the first two as space variables $t_1 = x$, $t_2 = y$, the third will
be $t_3 = t$. We keep the notation $\partial \equiv \partial_x \equiv
\partial_{t_1}$.

We can then write an infinite hierarchy of partial differential
equations, the Kadomtsev-Petviashvili hierarchy, in the following
compact form
\begin{equation}
\frac{\partial K}{\partial t_n} =	-( K\partial^n K^{-1})_- K.
\label{eq2.13}
\end{equation}

It follows from Theorem~\ref{thm1} that the flows (\ref{eq2.13}) all
commute, i.e.
\begin{equation}
[\partial_n, \partial_m]K = 0, \quad \partial_n = \partial \big/
\partial_{t_n}.
\label{eq2.14}
\end{equation}
Each operator equation (\ref{eq2.13}) (for a fixed $n \in \Bbb Z^>$)
gives rise to an infinite coupled set of PDEs for the coefficients
$K_n$ of eq.~(\ref{eq2.4}).

The equations all have the form
\begin{equation}
\partial_n K_j - \sum_{\ell = 1}^n \left({n \atop \ell-1}\right)
K_\ell^{(n - \ell +1)} =
\Pol(K_1, K_2 , \dots , K_{n+j-2}), \quad j = 1,2,3,\dots
\label{eq2.15}
\end{equation}
where $\Pol$ is a differential polynomial in $K_i$ and their
$x$-derivatives upto order $n-1$. The linear part of the polynomials
has been separated out in the left hand side of eq.~(\ref{eq2.15}), so
all terms on the right hand side are quadratic or higher. Note that
each equation in the system (\ref{eq2.15}) involves just one time
$t_n$ and the variable $x$.

A closed finite system of compatible equations involving $K_1, \dots,
K_n$ as unknowns and the independent variables $t_1 = x$, $t_2 = y,
\dots , t_n$ is obtained by taking the set of equations (\ref{eq2.15})
with labels $(n', j)$ satisfying
\begin{equation}
2 \le n' \le n, \quad 1 \le j \le n - n' + 1.
\label{eq2.16}
\end{equation}

For example, to obtain the Kadomtsev-Petviashvili equation itself,
take $n=3$ and hence $(n', j) = (3,1)$, $(2,1)$ and $(2,2)$. The
corresponding equations are
\begin{gather}
\partial_3 K_1 - (K_1^{'''} + 3K_2^{''} + 3K_3')	= 3K_1^2 K_1'
- 3K_1K_1^{''} - 3K_1^{\prime 2} - 3K_2' K_1 - 3K_1'K_2
\label{eq2.17}\\
\partial_2K_1 - (K_1^{''} + 2K_2') = -2K_1K_1'
\label{eq2.18}\\
\partial_2K_2 - (K_2^{''} + 2K_3')	= -2K_1'K_2.
\label{eq2.19}
\end{gather}
All three are evolution equations (in $t_3$, $t_2$ and $t_2$,
respectively) for the functions $K_1$, $K_2$ and $K_3$. Using
(\ref{eq2.19}) and (\ref{eq2.18})
to eliminate $K_3$ and $K_2$ from eq.~(\ref{eq2.17}), we obtain the
potential KP equation
\begin{equation}
w_{xt}	= \frac{1}{4} w_{xxxx} + \frac{3}{2} w_x w_{xx} + \frac{3}{4} w_{yy}
\label{eq2.20}
\end{equation}
for the variable
\begin{equation}
w(x,y,t)	= -2K_1(t_1,t_2,t_3).
\label{eq2.21}
\end{equation}
(the usual KP equation is satisfied by $u \equiv w_x$).

A further example is obtained by taking $n=4$ and hence $(n';j) =
(4;1)$, $(3;1)$, $(3;2)$, $(2;1)$, $(2;2)$, $(2;3)$. In addition to
eq.~(\ref{eq2.17}), (\ref{eq2.18}) and (\ref{eq2.19}) we obtain 3
further evolution equations:
\begin{multline}
\partial_4K_1 - (K_1^{''''} + 4K_2^{'''} + 6K_3^{''} + 4K_4')
= -10K_1' K_1^{''} + 12K_1 K_1^{\prime 2} - 4K_1^3 K_1' + 6K_1^2 K_1^{''}\\
- 4K_1 K_1^{'''} - 12K_1' K_2' + 4K_1^2 K_2'	+ 8K_1K_1'K_2 -
6K_1^{''}K_2 - 6K_1K_2^{''}
- 4K_2 K_2' - 4K_1'K_3 - 4K_1K_3'
\label{eq2.22}
\end{multline}
\begin{gather}
\partial_3K_2 - (K_2^{'''} + 3K_3^{''} + 3K_4') =
	-3K_1' K_2' + 3K_1 K_1'K_2 - 3K_1^{''} K_2 - 3K_2 K_2' - 3K_1' K_3
\label{eq2.23}\\
\partial_2 K_3 - (K_3^{''} + 2K_4') = -2K_1' K_3.
\label{eq2.24}
\end{gather}

We can consider eq.~(\ref{eq2.22}) to be the basic evolution equation
for $K_1$ and eliminate $K_2$, $K_3$ and $K_4$ using the other
equations. The result is not unique. It leads for instance to the
second equation of what Jimbo and Miwa~\cite{ref7} call the
KP-hierarchy (the first higher order KP equation):
\begin{equation}
w_{xxxy} + 3w_{xy} w_x + w_y w_{xx} + 2w_{yt} - 3w_{xz} = 0 \quad
\bigl( w = -2K_1(t_1 = x, t_2 = y, t_3 = t, t_4 = z) \bigr).
\label{eq2.25}
\end{equation}
Eq.~(\ref{eq2.25}) was called the Jimbo-Miwa equation in
Ref.~\cite{ref24,ref25}.

Alternatively, the term $w_y$ can be eliminated, using
eq.~(\ref{eq2.25}) together with the KP itself (eq.~(\ref{eq2.20})),
to obtain the usual higher order KP equation~\cite{ref8} involving $x,
y$ and $z$ only:
\begin{equation}
w_{xxxxy} + 6w_{xx} w_{xy} + 4w_x w_{xxy} + 2w_{xxx} w_y - 2w_{xxz} +
w_{yyy} = 0.
\label{eq2.26}
\end{equation}

Notice that in the form of eq.~(\ref{eq2.13}) (or (\ref{eq2.15})) we
have local evolution equations, whereas the KP equation (\ref{eq2.20})
and the JM equation (\ref{eq2.25}) viewed as evolution equations for
$w$ in $t$ or $z$, respectively, are nonlocal.

Finally we remark that the entire KP hierarchy can be written in terms
of the function $w = -2K_1(t_1, t_2, t_3, \dots)$ as
\begin{equation}
\frac{\partial w}{\partial t_n} = 2 \res_\partial K\partial^n K^{-1},
\quad n = 1,2,\dots
\label{eq2.27}
\end{equation}
where all higher terms $K_2$, $K_3, \dots$ are excluded using
\begin{equation}
K'_{j+1} = \frac{1}{2} \partial_2 K_j + K_1' K_j - \frac{1}{2} K_j^{''}.
\label{eq2.28}
\end{equation}
Eq.~(\ref{eq2.28}) follows from eq.~(\ref{eq2.13}) for $n=2$.

For each $n$ the right hand side of eq.~(\ref{eq2.27}) involves only
$w$, its $y$-derivatives upto order $n-1$ and a finite number of
$x$-derivatives and $x$ integrals of $w$. Thus we have
\begin{equation}
\begin{gathered}
\frac{\partial w}{\partial t_1} = w' = w_x\\
\frac{\partial w}{\partial t_2} = w_y\\
\frac{\partial w}{\partial t_3} = \frac{1}{4} w_{xxx} + \frac{3}{4}
w_x^2 + \frac{3}{4} \partial^{-1} w_{yy}\\
\frac{\partial w}{\partial t_4} = \frac{1}{2} w_{xxxy} + 4w_xw_y - 2
\partial^{-1} w_{xx} w_y + \partial^{-2} w_{yyy}.
\end{gathered}
\label{eq2.29}
\end{equation}

\subsection{Formal Baker-Akhiezer functions}

Below we shall make use of the formal Baker-Akhiezer
functions~\cite{ref8} (first introduced by G. Wilson~\cite{ref26'}.
They can be defined in terms of the PDO $K$ of eq.~(\ref{eq2.4}) by putting
\begin{equation}
\begin{gathered}
\varphi(\lambda)	= Ke^\zeta = e^\zeta \biggl( 1 + \sum_{n=1}
\lambda^{-n} K_n (\vec t) \biggr)\\
\varphi^*(\lambda) = K^{*-1} e^{-\zeta} = e^{-\zeta} \biggl( 1 +
\sum_{n=1} \lambda^{-n} K_N^*(\vec t) \biggr)\\
\zeta = \sum_{k=1} \lambda^k t_k, \quad \vec t = t_1, t_2, \dots,
\quad t_1 = x.
\end{gathered}
\label{eq2.30}
\end{equation}
We shall need the following lemma, due to Dickey~\cite{ref8}.

\begin{lem} Let $P = \sum_k p^k \partial^k$ and $Q = \sum_k q^k
\partial^{-k}$ be any two PDOs and let
\begin{equation}
\res_{\partial} \Sigma a_n \partial^{-n} = a_{1}.
\label{eq2.31}
\end{equation}
Then we have
\begin{equation}
\res_{\lambda} (Pe^{\lambda x}) (Qe^{-\lambda x}) = \res_\partial PQ^*.
\label{eq2.32}
\end{equation}
\end{lem}

\begin{pf} A direct calculation of the two sides of eq.~(\ref{eq2.30})
shows that both are equal to:
$$
\lhs = \rhs = \sum_{n+m=1} (-1)^m p_n q_m.
$$
We have used the fact that the formula $\partial^n e^{\lambda x} =
\lambda^n e^{\lambda x}$ holds for $n$ positive and negative in a
formal calculus.\cite{ref8}

Using the above lemma, eq.~(\ref{eq2.27}) and the definition
(\ref{eq2.28}) we obtain a formula summing up the KP hierarchy in
terms of the formal Baker-Akhiezer functions:
\begin{equation}
\frac{\partial w}{\partial t_m}	= 2 \res_\lambda \lambda^m \varphi \varphi^*,
\label{eq2.33}
\end{equation}
which can be rewritten as
\begin{equation}
2 \varphi \varphi^* =	2 + \frac{w_x}{\lambda^2} +
\frac{w_y}{\lambda^3} + \frac{w_t}{\lambda^4} + \dots
\label{eq2.34}
\end{equation}
The asterisk in eq.~(\ref{eq2.34}) has the same meaning as in
eq.~(\ref{eq2.7}), i.e. it does not indicate conjugation.
The term $\lambda^{-1}$ is absent in eq.~(\ref{eq2.34}), since we have
$K_1 + K_1^* = 0$. The $\lambda^0$ term on the right hand side of
eq.~(\ref{eq2.34}) follows from the asymptotic behavior of the Baker
functions for $\lambda \to + \infty$.
\end{pf}

\subsection{Symmetries of the KP hierarchy}

We shall call a ``symmetry'' of the KP equation any differential, or
integrodifferential equation of the form
\begin{equation}
\frac{\partial w}{\partial t} = F(w, t),
\label{eq2.35}
\end{equation}
where $F$ is a function of $w$, its $x$ and $y$ derivatives and of
integrals of the type $\partial^{-k} w$, $k \in \Bbb Z^>$, that is
compatible with the KP equation itself. Similarly, a symmetry of the
KP hierarchy will be any equation of the form (\ref{eq2.35}),
compatible with the entire hierarchy.

With this interpretation, all higher equations in the KP hierarchy are
symmetries of the KP equation itself. Since all of the corresponding
flows (see eq.~(\ref{eq2.14})) commute, these symmetries generate an
abelian Lie algebra.

Other symmetries of the KP equation exist and, contrary to the KP
hierarchy, typically the corresponding equations (\ref{eq2.35}) have
coefficients depending on the independent variables $x$, $y$, $t_n$.
Among them we mention the Lie points
symmetries~\cite{ref17,ref18,ref22,ref23} and also some higher
symmetries~\cite{ref14}. The most complete treatment of all
symmetries of the KP hierarchy in the framework of integrability
theory is given in Ref.~\cite{ref9},\dots ,~\cite{ref12}. To put
those results into the present context, let us use the mapping of
Section~II.B from the space of pseudodifferential operators to vecctor
fields.

We choose the PDOs to be
\begin{equation}
C_{mn} = \hat x^n \partial^m, \quad \hat x = \sum_{k=1}^\infty k t_k
\partial^{k-1}, \qquad n,m \in \Bbb Z
\label{eq2.36}
\end{equation}
and construct the mapping $C_{mn} \to V_{mn}$ with
\begin{equation}
V_{mn} K = [\partial_{mn}, K] = -(K\hat x^n \partial^m K^{-1})_- K.
\label{eq2.37}
\end{equation}
It follows from Theorem~\ref{thm1} that we have
\begin{equation}
(\partial_{mn} \partial_k - \partial_k \partial_{mn})K = 0
\label{eq2.38}
\end{equation}
and hence for any values of $m$ and $n$ the flow of (\ref{eq2.36}) is
compatible with the flows of the KP hierarchy.

The Lie algebra of the vector fields $V_{mn}$, generated by $\hat x^n
\partial_n^m$ for $n \ne 0$, $n \in \Bbb Z$ and $\partial_{t_m} -
\partial_x^m$ for $n=0$, is the algebra of ``additional symmetries''
of the KP hierarchy (they have also been called the ``Orlov-Schulman
symmetries''). The commutation relations have the form:
\begin{equation}
\begin{gathered}
[V_{mn}, V_{m'n'}]	= [\partial_{mn}, \partial_{m'n'}] \sim [x^n
\partial^m, x^{n'} \partial^{m'}], \quad n \ne 0, n' \ne 0\\
[V_{mn}, V_{m'n'}]	= 0 \text{ for } n = 0, \text{ or } n' = 0.
\end{gathered}
\label{eq2.39}
\end{equation}

The algebra of vector fields $V_{mn}$ contains infinitely many
different sets of conformal Lie algebras as subalgebras (Virasoro
algebras without a central extension). Amongst them we mention two
that are of particular interest. The first corresponds to $v(x)
\partial$, where $v(x)$ is an arbitrary function. This Virasoro
algebra figures in Ref.~\cite{ref17} and~\cite{ref18} among the point
symmetries and corresponds to arbitrary reparametrizations of time.
The other one corresponds to $xv(\partial)$, was introduced in
Ref.~\cite{ref9}--\cite{ref12}, and corresponds to a reparametrization
of the spectral parameter. It has been used for establishing
relations between soliton theory and quantum field
theory~\cite{ref26,ref27}.

This algebra has infinitely many different infinite dimensional
abelian subalgebras. Any one of them can be chosen to generate a
hierarchy of commuting flows. One of them is the well known KP
hierarchy itself. The others could be called ``alternative KP
hierarchies''. As in the case of the KP hierarchy, it is possible to
construct finite closed subsystems of partial differential equations
for a subset of the functions $K_i$. As in Section~II.C it is then
possible to construct evolution equations (in general nonlocal ones)
for one function $w$, evolving in a 3-dimensional space, spanned by
$x$, $\tau_1$ and $\tau_2$ where $\partial_{\tau_1}$ and
$\partial_{\tau_2}$ are any independent commuting combinations of the
operators $\partial_{mn}$ of eq.~(\ref{eq2.37}).

The flow of the function $w = -2K_1$ with respect to the ``time''
$t_{mn}$ of eq.~(\ref{eq2.37}) is given by the following theorem.

\begin{thm} \label{thm2}
Let $K$ be the PDO of eq.~(\ref{eq2.4}) and $t_{mn}$ the time defined
in eq.~{\em(\ref{eq2.35})} and {\em(\ref{eq2.36})},
{\em(\ref{eq2.37})}. The flow of $w = -2K_1$ with respect to this
time is given as
\begin{equation}
\partial_{mn} w = 2 \res_\partial (K \hat x^n \partial^m K^{-1}),
\label{eq2.40}
\end{equation}
or equivalently
\begin{equation*}
\partial_{mn} w = 2 \res_\lambda \lambda^m \frac{\partial^n
\varphi(\lambda)}{\partial \lambda^n} \varphi^*(\lambda).
\end{equation*}

\end{thm}

\begin{pf} Theorem~\ref{thm2} is a consequence of the Lemma of
Section~II.D. We set $P = K \hat x \partial^m$, $Q = K^{*-1}$ and use
the definition (\ref{eq2.30}) of the formal Baker-Akhiezer functions.
We then have
\begin{equation}
\res_\partial K \hat x^n \partial^m K^{-1}
	= \res_\lambda(K \hat x^n \partial^m e^\zeta)(K^{*-1} e^{-\zeta})
	= \res_\lambda \lambda^m (K \hat x^n e^\zeta)(K^{*-1} e^{-\zeta})
	= \res_\lambda \lambda^m \frac{\partial^n
\varphi(\lambda)}{\partial \lambda^n} \varphi^*(\lambda).
\end{equation}
\end{pf}

\subsection{Symmetries via vertex operators}

In order to link the Jimbo-Miwa $\widehat{\gl}(\infty)$
symmetries~\cite{ref7} with those of Orlov and Schulman,\cite{ref9},
\dots ,~\cite{ref12} we need some results on vertex operators and
$\tau$-functions. The vertex operators can be written as
\begin{align}
X(\lambda)	&= \exp \biggl(t_0 \ln \lambda + \sum_{k=1}^\infty
\lambda^k t_k \biggr) \exp \biggl( -\partial_0 - \sum_{k=1}^\infty
\frac{1}{k \lambda^k} \partial_k \biggr)
\label{eq2.41}\\
X^*(\lambda) &= \exp \biggl( -t_0 \ln \lambda - \sum_{k=1}^\infty
\lambda^k t_k \biggr) \exp \biggl( \partial_0 + \sum_{k=1}^\infty
\frac{1}{k \lambda^k} \partial_k \biggr).
\label{eq2.42}
\end{align}
The zero mode $t_0 \ln \lambda - \partial_0$ was added to the vertex
operator in Ref.~\cite{ref28} to simplify calculations. It can be
checked by a direct calculation that the vertex operators introduced
above satisfy the fermion algebras anticommutation relations:
\begin{equation}
\begin{gathered}
X(\lambda) X(\mu) + X(\mu) X(\lambda) = 0,\\
X^*(\lambda) X^*(\mu) + X^*(\mu) X^*(\lambda) = 0,\\
X(\mu) X^*(\lambda) + X^*(\lambda) X(\mu) = \delta (\lambda - \mu),
\end{gathered}
\label{eq2.43}
\end{equation}
where $\delta(\lambda - \mu)$ is the Dirac $\delta$-function with
respect to integration about a circle $S^1$ (close to $\lambda \to \infty$)
\begin{equation}
\begin{gathered}
\delta(\lambda - \mu) = \sum_{n=-\infty}^\infty \biggl(
\frac{\mu}{\lambda} \biggr)^n \frac{1}{\lambda}\\
\oint f(\lambda) \delta(\lambda - \mu)d \lambda = f(\mu)
\end{gathered}
\label{eq2.44}
\end{equation}
The ``zero-time'' $t_0$ is a discrete variable, introduced in
Ref.~\cite{ref7} and~\cite{ref29}, though it was not used to add a
zero mode to the vertex operators (\ref{eq2.41}) and (\ref{eq2.42}).
One of the uses it was put to in Ref.~\cite{ref28} was to introduce
the flag space of Grassmannians into KP theory.

A $\tau$-function is a function of all the times $\evec t = \{ t_1,
t_2, \dots \}$ and also of the discrete ``zero'' time
$t_0$~\cite{ref7,ref29}
\begin{equation}
\tau	= \tau_n (\vec t), \quad n = t_0.
\label{eq2.45}
\end{equation}

The Baker function near $\lambda = \infty$ can be expressed in terms
of vertex operators and the $\tau$-function as
\begin{equation}
\varphi(\lambda, t_0, \vec t) = \frac{X(\lambda, t_0, \vec t) \tau(t_0
+ 1, \vec t)}{\tau(t_0, \vec t)}
\label{eq2.46}
\end{equation}
(with a similar expression for $\varphi^*$).

For any sufficiently small shift $(\vec t - \vec t')$ we have the
bilinear identity
\begin{equation}
\res_{\lambda = \infty} \varphi( \lambda, t_0, \vec t)
\varphi^*(\lambda, t_0, \vec t') = 0.
\label{eq2.47}
\end{equation}

Let us now consider variations of the $\tau$-function due to the vector field
\begin{equation}
V_{\lambda \mu} \tau	= X(\lambda) X^*(\mu) \tau.
\label{eq2.48}
\end{equation}

In the Jimbo-Miwa approach~\cite{ref7} this is an infinitesimal group
transformation of the $\tau$-function, corresponding to an action of
the algebra $\widehat \gl(\infty)$. Let us calculate the
corresponding induced action on KP solutions $w_1$, following
ref.~\cite{ref9}. We have
\begin{equation}
V_{\lambda \mu} w	= -2 \res_{k=\infty} \bigl(V_{\lambda \mu}
\varphi(k) \bigr) \varphi^*(k),
\label{eq2.49}
\end{equation}
where $k$ is a spectral parameter and all times in $\varphi(k)$ and
$\varphi^*(k)$ are set equal. In deriving (\ref{eq2.49}) use is made
of the relations
\begin{equation}
\varphi = e^\zeta \Biggl( 1 - \frac{w}{2\lambda} + O \biggl(
\frac{1}{\lambda^2}\biggr)\Biggr),
\quad \varphi^* = e^{-\zeta} \Biggl( 1 + O \biggl( \frac{1}{\lambda}
\biggr)\Biggr).
\label{eq2.50}
\end{equation}
We recall that the relation between the $\tau$-functions and solutions is
\begin{equation}
w	= 2 \frac{\partial}{\partial x} \ln \tau,
\label{eq2.51}
\end{equation}
but we do not use that here. For the Baker function we have
\begin{equation}
V_{\lambda \mu} \varphi(k)	= \frac{X(k)(V_{\lambda\mu}
\tau)}{\tau} - \varphi(k) \frac{V_{\lambda \mu} \tau}{\tau}.
\label{eq2.52}
\end{equation}
The last term in eq.~(\ref{eq2.52}) does not contribute to the residue
in eq.~(\ref{eq2.49}) and we obtain
\begin{equation}
V_{\lambda\mu}w	= -2 \res_k \frac{\bigl(X(k)
X(\lambda)X^*(\mu)\tau\bigr)}{\tau} \varphi^*(k).
\label{eq2.53}
\end{equation}

{}From the fermion commutation relations (\ref{eq2.43}) we obtain
\begin{equation}
\bigl[X(k), X(\lambda)X^*(\mu)\bigr] = -\delta(k - \mu) X(\lambda)
\label{eq2.54}
\end{equation}
and hence
\begin{equation}
- \res_k \frac{\bigl(X(k) X(\lambda) X^*(\mu)\tau\bigr)}{\tau}
\varphi^*(k) = \res_k \delta(k - \mu) \frac{X(\lambda)\tau}{\tau}
\varphi^*(k) - \res_k \frac{X(\lambda) X^*(\mu) X(k)\tau}{\tau} \varphi^*(k).
\label{eq2.55}
\end{equation}

Making use of the bilinear identity (\ref{eq2.47}) we can see that the
last term in eq.~(\ref{eq2.55}) vanishes.

Using eq.~(\ref{eq2.53}), (\ref{eq2.54}) and (\ref{eq2.46}) we obtain
the final result
\begin{equation}
V_{\lambda\mu} w = 2 \varphi(\lambda) \varphi^*(\mu).
\label{eq2.56}
\end{equation}
This formula was announced in Ref.~\cite{ref12} (with unfortunate
misprints) and proven in Ref.~\cite{ref9}. Eq.~(\ref{eq2.56}) plays a
key role if we wish to relate the symmetries (\ref{eq2.40}) with the
$\widehat{\gl}(\infty)$ transforms~\cite{ref7}. Using
eq.~(\ref{eq2.48}) and (\ref{eq2.56}) we can write the following
commutative diagram
\begin{equation}
\begin{array}{ccc}
V_{\lambda\mu} \tau = X(\lambda) X^*(\mu)\tau &\longrightarrow&
V_{\lambda\mu} w = 2\varphi(\lambda) \varphi^*(\mu)\\
\downarrow && \downarrow\\
\partial_{mn}\tau =\res \lambda^m \frac{\partial^n
X(\lambda)}{\partial \lambda^n} X^*(\lambda)\tau &\longrightarrow&
\partial_{mn}w = 2 \res \lambda^m \frac{\partial^n\varphi}{\partial
\lambda^n} \varphi^*.
\end{array}
\label{eq2.58}
\end{equation}

Below, in Section IV, we shall make use of the formalism of
Sections~II.E and II.F to obtain all point symmetries of the equations
of the KP-hierarchy.

\section{Point symmetries and conditional symmetries o the KP and JM equations}

The Lie point symmetries of the potential KP equation (\ref{eq2.20})
were found in Ref.~\cite{ref18}, using a standard
algorithm~\cite{ref15,ref16}, that does not depend on the
integrability of the equation. Indeed, what was found was the Lie
algebra of vector fields of the form
\begin{equation}
\hat V = \xi \partial_x + \eta \partial_y + \tau \partial_t + \phi \partial_w,
\label{eq3.1}
\end{equation}
where $\xi$, $\eta$, $\tau$ and $\phi$ are functions of $x$, $y$, $t$,
and $w$, such that the fourth prolongation $\pr^{(4)}V$ annihilates
the studied equation (\ref{eq2.20}) on its solution set. Once
integrated, the Lie algebra provides the group transformations, taking
solutions into solutions.

The general element of the symmetry algebra of the potential KP
equation (\ref{eq2.20}) was shown to have the form
\begin{equation}
\hat V = T(f) + Y(g) + X(h) + W(k) + U(\ell)
\label{eq3.2}
\end{equation}
with
\begin{gather*}
T(f)	= f\partial_t + \frac23 yf'\partial_y + \frac13 \biggl[xf' +
\frac23 y^2f{''} \biggr]\partial_x
- \biggl[ \frac13 wf' + \frac19 x^2f{''} + \frac4{27} xy^2 f{'''} +
\frac{4}{243} y^4 f^{''''} \biggr]\partial_w,\\
Y(g) = g\partial_y + \frac23 yg' \partial_x - \frac49y \biggl( xg{''}
+ \frac29 y^2 g{'''} \biggr) \partial_w,\\
X(h)	= h \partial_x - \frac23 \biggl( xh' + \frac23 y^2h{''}
\biggr)\partial_w,\\
W(k) = ky \partial_w,\\
U(\ell) = \ell \partial_w,
\end{gather*}
where $f$, $g$, $h$, $k$ and $\ell$ are functions of time $t$ (the
sign of time $t$ has been changed with respect to Ref.~\cite{ref18}).

The ``symmetry'' in the sense of eq.~(\ref{eq2.35}), i.e. the equation
compatible with the PKP equation (\ref{eq2.20}) in this case is a
first order linear equation, namely
\begin{multline}
fw_t + \biggl( \frac23 yf' + g \biggr) w_y + \biggr[ \frac13 xf' +
\frac29 y^2f{''} + \frac23 yg' + h \biggr]w_x \\
+ \biggl[ \frac13 wf' + \frac19 x^2f{''} + \frac4{27} xy^2 f{'''} +
\frac{4}{243} y^4 f{''''} + \frac49 xyg{''} - \frac{8}{81}
y^2 g{'''} + \frac23 xh' - \frac49 y^2 h{''} - yk - \ell \biggr] = 0.
\label{eq3.3}
\end{multline}

The vector fields (\ref{eq3.2}) can of course be integrated to yield
point transformations, taking solutions of the PKP equation into solutions.

The nonzero commutation relations of the symmetry algebra are
\begin{equation}
\bigl[T(f_1), T(f_2)\bigr]
	= T(f_1f_2' - f_1'f_2)
\label{eq3.4}
\end{equation}
\begin{equation}
\begin{gathered}
\bigl[Y(g_1), Y(g_2)\bigr]	= \frac23 X(g_1g_2' - g_1'g_2)\\
\bigl[Y(g), X(h)\bigr] = \frac49W(hg{''} - g'h') - \frac89 U(gh{''})\\
\bigl[Y(g), W(k)\bigr] = U(gk)\\
\bigl[X(h_1), X(h_2)\bigl] = -\frac23 U(h_1h_2' - h_1'h_2)
\end{gathered}
\label{eq3.5}
\end{equation}
\begin{equation}
\begin{gathered}
\bigl[T(f), Y(g)\bigr] = Y\biggl(f\dot g - \frac23 g \dot f\biggr)\\
\bigl[T(f), X(h)\bigr] = X\biggl(f \dot h - \frac13 h \dot f\biggr)\\
\bigl[T(f), W(k)\bigr] = W(f \dot k + \dot f k).\\
\bigl[T(f), U(\ell)\bigr] = U\biggl(f \dot \ell + \frac13 \ell \dot f\biggr)
\end{gathered}
\label{eq3.6}
\end{equation}
We see that the fields $T(f)$ form a centerless Virasoro algebra (or
conformal Lie algebra). The vector fields $\{ Y(g), X(h), W(k),
U(\ell) \}$ form a centerless Kac-Moody algebra. Actually, this is a
loop algebra with $t$ as the loop parameter. The underlying Lie
algebra is a 9 dimensional nilpotent Lie algebra that can be imbedded
into $\Sl(8, \Bbb R)$. A basis is provided by the vector fields
\begin{equation}
\begin{gathered}
Y = X_{1,1} = \partial_y, \ Q = X_2 = y\partial_x, \ H = X_{3,1} = xy
\partial_w, X = X_{3,2} = \partial_x\\
U = X_{4,1} = x \partial_w, \quad W_{3} = X_{4,2} = y^3 \partial_w,
\quad W_2 = X_{5,1} = y^2 \partial_w, \quad W_1 = X_{6,1} = y
\partial_w, \quad W_0 = X_{7,1} = \partial_w.
\end{gathered}
\label{eq3.7}
\end{equation}
There is a natural grading on this algebra and the first label on
$X_{i,j}$ determines the place of each element $X_{i,j}$ in this
grading. The embedding into $\Sl(8, \Bbb R)$ is given by the matrices
\begin{equation} \left(
\begin{array}{cccccccc}
0	&0	&0	&-x	&0	&0	&-w_1	&w_0\\
0	&0	&0	&q	&0	&0	&2w_2	&-w_1\\
0	&0	&0	&0	&0	&0	&-6w_3	&2w_2\\
0	&0	&0	&0	&0	&0	&h	&-u\\
0	&0	&0	&0	&0	&0	&q	&-x\\
0	&0	&0	&0	&0	&0	&0	&0\\
0	&0	&0	&0	&0	&0	&0	&y\\
0	&0	&0	&0	&0	&0	&0	&0\\
\end{array}
\right).
\label{eq3.8}
\end{equation}

The Jimbo-Miwa equation (\ref{eq2.25}), taken on its own, has been
shown not to pass any of the usual integrability
criteria~\cite{ref24}. Moreover, its symmetry algebra does not have a
Kac-Moody-Virasoro structure~\cite{ref24}. An alternative approach
was taken in Ref.~\cite{ref25}, namely that of ``conditional
symmetries''. Thus, Lie point symmetries were found that leave only a
subset of the solution set of the JM equation invariant, namely those
that simultaneously satisfy the PKP equation (\ref{eq2.20}).

The corresponding Lie algebra does have a Kac-Moody-Virasoro structure
and its general element can be written as~\cite{ref25}
\begin{equation}
V	= Z(f) + T(g) + Y(h) + X(k) + W(g)
\label{eq3.9}
\end{equation}
\begin{multline}
Z(f)	= f\partial_z + \frac34 t \dot f \partial_t + \frac{1}{128}
[32 x \dot f + 48 yt \ddot f + 9 t^3 \dddot f]\partial_x\\
+ \frac{1}{32} [16 y \dot f + 9 t^2 \ddot f] \partial_y -
\frac{1}{128} [32w \dot f + 32xy \ddot f + 6t(4y^2 + 3tx) \dddot f
+ 9 yt^3 f^{i v}] \partial_w
\label{eq3.10}
\end{multline}
\begin{equation}
\begin{gathered}
T(g)	= g \partial_t + \frac{1}{32} [16y \dot g + 9 t^2 \ddot
g]\partial_x + \frac34 t \dot g \partial_y -
\frac{1}{3} \bigl[4(3tx + 2y^2) \ddot g + 9y t^2 \dddot g\bigr]\partial_w\\
Y(h)	= h \partial_y + \frac34 t \dot h \partial_x - \frac{1}{4} [2x
\dot h + 3ty \ddot h] \partial_w\\
X(k) = k \partial_x - y \dot k \partial_w\\
W(G)	= G(z,t) \partial_w
\end{gathered}
\label{eq3.11}
\end{equation}
where $f$, $g$, $h$ and $k$ are functions of $z$.

Finally we note that the symmetry algebra of the first higher order KP
equation (\ref{eq2.24}) can either be calculated directly, or can be
obtained from the ``conditional'' symmetry algebra (\ref{eq3.10}),
(\ref{eq3.11}). Indeed, eq.~(\ref{eq2.24}) does not involve the
variable $t$, neither in coefficients, nor in derivatives. Hence, we
can consider $t$ to be constant in eq.~(\ref{eq3.10}). This amounts
to dropping the Lie algebra elements $T(g)$, after using them to
eliminate the $t$-derivative in $Z(f)$.

The result is again a Kac-Moody-Virasoro algebra as the symmetry
algebra of eq.~(\ref{eq2.26}), namely
\begin{equation}
V	= Z(f) + Y(h) + X(k) + W(g),
\label{eq3.11'}
\end{equation}
with
\begin{equation}
\begin{gathered}
Z(f)	= f \partial_z + \frac14 \dot f(x \partial_x + 2y \partial_y)
- \frac14 (w \dot f + xy \ddot f) \partial_w,\\
Y(g)	= g\partial_y - \frac12 x \dot g \partial_w,\\
X(h)	= h\partial_x - y \dot h \partial_w,\\
W(k)	= k\partial_w,
\end{gathered}
\label{eq3.12}
\end{equation}
where $f(z)$, $g(z)$, $h(z)$ and $k(z)$ are all arbitrary functions of
the time $t_4 = z$.

Exactly the same symmetry algebra (\ref{eq3.12}) is obtained by using
a standard algorithm, presented e.g. in Ref.~\cite{ref15} and
implemented in a MACSYMA program~\cite{ref16}. The Virasoro
subalgebra of eq.~(\ref{eq3.12}) corresponds to the vector fields
$Z(f)$. The remaining vector fields $\bigl\{ Y(h), X(k), W(g)
\bigr\}$ form a centerless Kac-Moody algebra. This is a loop algebra
with $z$ as the loop parameter. The underlying Lie algebra is a
five-dimensional nilpotent Lie algebra with the basis
\begin{equation}
X = \partial_x, \quad Y = \partial_y, \quad P_1 = x\partial_w, \quad
P_2 = y \partial_w, \quad H = \partial_w,
\label{eq3.13}
\end{equation}
isomorphic to the Heisenberg algebra in two dimensions. The algebra
(\ref{eq3.13}) can be imbedded into $\Sl(4, \Bbb R)$ as follows
\begin{equation}
M = \left(
\begin{array}{cccc}
0	&p_1	&p_2	&h\\
0	&0	&0	&x\\
0	&0	&0	&y\\
0	&0	&0	&0
\end{array}
\right).
\label{eq3.14'}
\end{equation}

An earlier observation is that all known integrable PDEs in
3-dimensions have Kac-Moody-Virasoro algebras as Lie point symmetry
algebras. To the KP, equation~\cite{ref17}, the PKP
equation~\cite{ref18}, the Davey-Stewartson equation~\cite{ref19}, the
3 wave resonant interaction equations~\cite{ref20} and several others,
we have just added the higher order KP equation (\ref{eq2.26}).

In Section~IV we shall show that the same is true for the entire KP hierarchy.

\section{Lie point symmetries of the KP hierarchy from the $\gl(\infty)$ ones}
\subsection{Extraction of local symmetries}

Let us now present the main result of this article, namely show that
the Lie point symmetry algebra of each equation in the KP hierarchy
has a Kac-Moody-Virasoro structure. The corresponding Lie point
symmetries can be directly extracted from the symmetries generated by
the pseudodifferential operators $C_{mn}$ of eq.~(\ref{eq2.36}) via
Theorem~\ref{thm2}. Moreover, we will show that all the symmetries
given in eq.~(\ref{eq2.40}) that are local (no integrals), are point
symmetries.

To see this explicitly let us rewrite the symmetries (\ref{eq2.40}) of
the KP hierarchy using a different basis. Instead of the operators
$C_{mn} = \hat x^n \partial^m$ of eq.~(\ref{eq2.36}), let us consider
an arbitrary function $h(x)$ (that can be expanded into a power
series, or Laurent series on the real line). We shall consider the
operators
\begin{equation}
h_{\alpha, N}	= \lambda^\alpha h \biggl( \frac{1}{N \lambda^{N-1}}
\frac{\partial}{\partial\lambda} \biggr).
\label{eq4.1}
\end{equation}
We have
\begin{equation}
h \biggl( \frac{1}{N\lambda^{N-1}} \frac{\partial}{\partial \lambda}
\biggr) = h
\biggl( \frac{\partial}{\partial E}\biggr) = \sum_n h_n \partial_E^n,
\qquad E = \lambda^N, \quad h_n = \const.
\label{eq4.2}
\end{equation}
We rewrite the symmetries (\ref{eq2.40}) in the form
\begin{equation}
V^N (\alpha, h)w	= 2 \res_\lambda \lambda^\alpha \biggl[ h
\biggl( \frac{1}{N\lambda^{N-1}} \frac{\partial}{\partial \lambda}
\biggr) \varphi(\lambda) \biggr] \varphi^* (\lambda)
\label{eq4.3}
\end{equation}
where $V^N(\alpha, h)$ is a vector field acting on $w$ (a linear
combination of the ``time derivatives'' $\partial_{mn}$).

For $h(x) = (Nx)^n$, $\alpha = m + nN - n$, we recover the symmetries
(\ref{eq2.40}) with $V^N(\alpha, h) \equiv \partial_{mn}$. In
particular, if we put $n=0$, we obtain the KP hierarchy itself namely
eq.~(\ref{eq2.33}).

Let us rewrite eq.~(\ref{eq4.3}) in a more convenient form, using
eq.~(\ref{eq2.30}) for the formal Baker-Akhiezer functions. Our aim
is to replace the power series in derivatives $\partial_\lambda$,
implicit in eq.~(\ref{eq4.3}), by power series in $\lambda$ itself.
The formula we are aiming at is
\begin{equation}
V^N(\alpha, h) w	= 2 \res \lambda^\alpha f(\lambda)
\varphi(\lambda) \varphi^* (\lambda),
\label{eq4.4}
\end{equation}
where $f(\lambda)$ is a function to be determined. We shall use the
following relation, valid both for differential and pseudodifferential
operators
\begin{equation}
f(\partial)e^\chi	= e^\chi f(\partial + \chi') \cdot 1,
\end{equation}
where $\chi' = \partial\chi/\partial x$.

\noindent Comparing eq.~(\ref{eq4.3}) and eq.~(\ref{eq4.4}) and using
eq.~(\ref{eq2.30}) we have
\begin{equation}
\begin{gathered}
f(\lambda)	= \varphi^{-1} (\lambda) h \biggl(
\frac{1}{N\lambda^{N-1}} \frac{\partial}{\partial\lambda} \biggr) e^\Phi,\\
\Phi	= \sum_{k=1}^\infty \lambda^k t_k + \ln \biggl( 1 + \sum_{n=1}
\lambda^{-n} K_n(t) \biggr).
\end{gathered}
\label{eq4.6}
\end{equation}
The KP hierarchy is written in terms of the function $w = -2K_1$. All
higher coefficients $K_n$ in eq.~(\ref{eq4.6}) are expressed
nonlocally in terms of $K_1$ (see eq.~(\ref{eq2.15}), \dots
,(\ref{eq2.24})). Since we will be using eq.~(\ref{eq4.4}) to extract
{\em local} symmetries, we shall drop all negative powers
$\lambda^{-n}$ in eq.~(\ref{eq4.6}), except for $\lambda^{-1}$. We obtain
\begin{equation}
f(\lambda)	= h \biggl\{ \frac{1}{N\lambda^{N-1}}
\frac{\partial}{\partial\lambda} + \sum_{k=1}^\infty \frac{k}{N}
\lambda^{k-N} t_k + \frac{w}{2N\lambda^{N+1}} + O \biggl(
\frac{1}{\lambda^{N+2}} \biggr) \biggr\} \cdot 1.
\label{eq4.7}
\end{equation}
Another ingredient in eq.~(\ref{eq4.4}) is the expansion
(\ref{eq2.34}) which we rewrite as
\begin{equation}
2 \varphi \varphi^* =	2 + \frac{w_x}{\lambda^2} +
\frac{w_y}{\lambda^3} + \frac{w_t}{\lambda^4} + \dots +
\frac{w_{t_n}}{\lambda^{N+1}} + O \biggl( \frac{1}{\lambda^{N+2}} \biggr).
\label{eq4.8}
\end{equation}
Before expanding the function $h\{ \quad \}$ into a Taylor series we
note that nonlocal terms (integrals of $w$) will be avoided only if
the terms $O(\lambda^{-N-2})$ do not participate in the calculation of
the residue in eq.~(\ref{eq4.4}). This imposes the necessary restriction
\begin{equation}
\alpha =	N, N-1, N-2, \dots , 2,1,0,-1.
\label{eq4.9}
\end{equation}
For the same reason, we must ``freeze'' all times in eq.~(\ref{eq4.7})
except $t_1 = x$, $t_2 = y$ and $t_N$, where $N$ is the number of the
equation in the KP hierarchy that we are considering ($N=3$ for the
PKP, $N=4$ for eq.~(\ref{eq2.26})). Thus, we set
\begin{equation}
t_k = 0, \quad k \ge N+1, \quad 3 \le k \le N-1
\label{eq4.10}
\end{equation}
in eq.~(\ref{eq4.7}).

Finally, we expand $f(\lambda)$ in a Taylor series about $t_N$, keep
only local terms and obtain
\begin{equation}
\begin{split}
V^N(\alpha h)w = \res \lambda^\alpha & \biggl\{ h(t_N) + h'(t_N)
\biggl[ \frac{2}{N\lambda^{N-2}} y +
\frac{1}{N\lambda^{N-1}}x + \frac{1}{2N\lambda^{N+1}}w \biggr]\\
&+ \frac{1}{2!} h{''}(t_N) \biggl[ \frac{4}{N^2\lambda^{2N-4}} y^2 +
\frac{4}{N^2\lambda^{2N-3}} xy + \frac{1}{N^2\lambda^{2N-2}}
(x^2 + 4y - 2Ny) \biggr]\\
&+ \frac{1}{3!} h{'''} (t_N) \biggl[ \frac{8y^3}{N^3 \lambda^{3N-6}} +
\frac{12xy^2}{N^3 \lambda^{3N-5}} \biggr]\\
&+ \frac{1}{4!} h{''''}(t_N) \frac{16y^4}{N^4 \lambda^{4N-8}} +
O(\lambda^{-N-2}) \biggr\} \biggl\{ 2 + \frac{w_x}{\lambda^2}
+ \frac{w_y}{\lambda^3} + \dots + \frac{w_{t_N}}{\lambda^{N+1}} + O
(\lambda^{-N-2}) \biggr\}
\end{split}
\label{eq4.11}
\end{equation}

All Lie point symmetries of all the equations of the KP hierarchy are
contained in eq.~(\ref{eq4.11}). To obtain them explicitly in terms
of vector fields of the form of eq.~(\ref{eq3.1}) with $t = t_3$
replaced by $t_N$ it will suffice to perform the substitutions
\begin{equation}
w \to -\partial_w, \quad w_x \to \partial_x, \quad w_y \to \partial_y,
\quad w_{t_N} \to \partial_{t_N}
\label{eq4.12}
\end{equation}
in eq.~(\ref{eq4.11})

\subsection{The Virasoro subalgebras of point symmetries}

Let us first of all obtain the Virasoro subalgebra of the symmetry
algebra for each equation in the KP hierarchy (throughout this article
we are dealing with a Virasoro algebra without a central charge). We
have in mind the equations written for the function $w(x,y,t,t_4,
\dots) = -2K_1(x,y,t,t_4, \dots)$ in which all variables except $x, y$
and $t_N$ are ``frozen'', i.e. all derivatives except those with
respect to $x, y$ and $t_N$ are eliminated from the $N$-th equation,
using the lower equations (see Section~II.C). To obtain the Virasoro
symmetries, we set $\alpha = N$ in eq.~(\ref{eq4.11}) and calculate
the residue. Let us consider each value of $N$ separately:
\begin{enumerate}
\item[1.] $N = 3$, $\alpha = 3$.

We obtain $T(f)$ as in eq.~(\ref{eq3.2})

\item[2.] $N = \alpha = 4$.

We obtain $Z(f)$ as in eq.~(\ref{eq3.12}), $f = f(t_4)$.

\item[3.] $N = \alpha = 5$.
\begin{equation}
T(f)	= f \partial_{t_5} + \frac{1}{5} f'(x\partial_x +
2y\partial_y) - \frac{1}{25} (5wf' + 4y^2f{''}) \partial_w, \quad f =
f(t_5).
\label{eq4.13}
\end{equation}

\item[4.] $N = \alpha \ge 6$.
\begin{equation}
T(f)	= f \partial_{t_N} + \frac{1}{N} f'(x\partial_x +
2y\partial_y) - \frac{1}{N} wf'\partial_w, \quad f = f(t_N).
\label{eq4.14}
\end{equation}
\end{enumerate}

The commutation relations are the same in all cases, namely those of
eq.~(\ref{eq3.4}).

\subsection{The Kac-Moody subalgebras of point symmetries} The
remaining local symmetries have a Kac-Moody structure in all cases and
are also obtained from eq.~(\ref{eq4.11}) by performing the
substitutions (\ref{eq4.12}) and calculating the residues for $\alpha
< N$ (see eq.~(\ref{eq4.9})). Again, the Kac-Moody algebras have no
central charge.

Specifically, we obtain the following results.

$N = 3$, $\alpha = 2,1,0$ and $-1$ yield the vector fields
(\ref{eq3.2}), \dots ,(\ref{eq3.2}), respectively.

$N=4$, $\alpha = 2,1$ and $-1$ yield the vector fields (\ref{eq3.12}),
(\ref{eq3.12}), (\ref{eq3.12}), respectively. The values $\alpha = 3$
and 0 lead to nonlocal symmetries.

$N = 5$, $\alpha = 2,1$ and $-1$ lead to
\begin{equation}
\begin{aligned}
Y(g) &= g\partial_y - \frac{4}{5} g'y\partial_w,\\
X(h)	&= h\partial_x,\\
W(k)	&= k\partial_w,
\end{aligned}
\label{eq4.15}
\end{equation}
respectively, where $g, h$ and $k$ are functions of $t_5$.

$N \ge 6$, $\alpha = 2, 1$ and $-1$ lead respectively to
\begin{equation}
\begin{aligned}
Y(g) &= g(t_N)\partial_y,\\
X(h)	&= h(t_N)\partial_x,\\
W(k)	&= k(t_N)\partial_w.
\end{aligned}
\label{eq4.16}
\end{equation}

We see that the Kac-Moody-Virasoro structure of the symmetry algebras
of the KP equation is no coincidence: it occurs for all equations of
the KP hierarchy. As $N$ increases, the specific character of the
Kac-Moody subalgebra simplifies. The underlying finite dimensional
nilpotent Lie algebra is a 9-dimensional subalgebra of $\Sl(8, \Bbb
R)$ for the KP itself (see eq.~(\ref{eq3.7}), (\ref{eq3.8})). For
$N=4$ it is a 5-dimensional subalgebra of $\Sl(4,\Bbb R)$, isomorphic
to the Heisenberg algebra $H(2)$ (see eq.~(\ref{eq3.13}),
(\ref{eq3.14'})). For $N=5$ this algebra is 4-dimensional, $\{
\partial_y, y\partial_w, \partial_w \} \oplus \{ \partial_x \}$ and
can be embedded into $\Sl(4,\Bbb R)$. For $N \ge 6$ the algebra is a
3-dimensional $\{ \partial_x, \partial_y, \partial_w \}$ and abelian.

\subsection{The symmetry groups}
To bring out further the role of the point symmetries, let us
construct the corresponding transformations by integrating the one
parameter subalgebras of the vector fields (\ref{eq3.1}), obtained
above
\begin{equation}
\begin{gathered}
\frac{d \tilde x}{d\lambda}
	= \xi(\tilde x, \tilde y, \tilde t, \tilde w), \quad \frac{d
\tilde y}{d\lambda}
	= \eta(\tilde x, \tilde y, \tilde t, \tilde w),\\
\frac{d \tilde t}{d\lambda}
	= \tau(\tilde x, \tilde y, \tilde t, \tilde w), \quad \frac{d
\tilde w}{d\lambda}
	= \phi(\tilde x, \tilde y, \tilde t, \tilde w), \quad
\tilde x \big|_{\lambda=0} = x, \quad \tilde y \big|_{\lambda=0} = y
\quad \tilde t \big|_{\lambda=0} = t, \quad \tilde w \big|_{\lambda=0} = w.
\end{gathered}
\label{eq4.17}
\end{equation}
\begin{description}
\item[$N \ge 6$]. Virasoro (eq.~(\ref{eq4.14})):
\begin{equation}
\tilde t_N	= \phi^{-1}\bigr(\lambda + \phi(t_N)\bigr), \quad
\tilde x	= x \biggl( \frac{f(\tilde t_N)}{f(t_N)}\biggr)^{1/N} , \quad
\tilde y	= y \biggl( \frac{f(\tilde t_N)}{f(t_N)}
\biggr)^{2/N}, \quad \tilde w	= w \biggl( \frac{f(\tilde
t_N)}{f(t_N)} \biggr)^{-1/N}
\label{eq4.18}
\end{equation}
Kac-Moody (eq.(\ref{eq4.16})):
\begin{equation*}
\tilde t = t, \quad \tilde x = x + \lambda h(t_N), \quad \tilde y = y
+ \lambda g(t_N), \quad \tilde w = w + \lambda k(t_N).
\end{equation*}
\item[$N = 5$]. Virasoro (eq.~(\ref{eq4.13})):
\begin{multline}
\tilde t_5 = \phi^{-1}\bigl(\lambda + \phi(t_5)\bigr), \quad \tilde x
= x \biggl( \frac{f(\tilde t_5)}{f(t_5)} \biggr), \\
\tilde y = y \biggl( \frac{f(\tilde t_5)}{f(t_5)} \biggr)^{2/5},\quad
\tilde w = \biggl( \frac{f(\tilde t_5)}{f(t_5)} \biggr)^{-1/5}
\biggl[ w - \frac{4}{25} y^2 \frac{\dot f(\tilde t) - \dot f(t)}{f(t)} \biggr]
\label{eq4.19}
\end{multline}
Kac-Moody (eq.(\ref{eq4.15}))
\begin{equation*}
\tilde t = t, \quad \tilde x = x + \lambda h(t_5), \quad \tilde y = y
+ \lambda g(t_5), \quad \tilde w = w - \frac{4}{5}
g'(t_5) y \lambda - \frac{2}{5} g(t_5) g'(t_5) \lambda^2
\end{equation*}
\item[$N=4$] Virasoro (eq.~(\ref{eq3.12})):
\begin{multline}
\tilde t_4 = \phi^{-1}\bigl(\lambda + \phi(t_4) \bigr), \quad	\tilde
x = x \biggl( \frac{f(\tilde t_4)}{f(t_4)}\biggr)^{1/4}, \\
\tilde y = y\biggl( \frac{f(\tilde t_4)}{f(t_4)}\biggr)^{1/2}, \quad
\tilde w =\biggl( \frac{f(\tilde t_4)}{f(t_4)}\biggr)^{-1/4}
\biggl( w - \frac{1}{4} xy \frac{\dot f(\tilde t_4) - \dot
f(t_4)}{f(t_4)} \biggr)
\label{eq4.20}
\end{multline}
Kac-Moody (eq.~(\ref{eq3.12}), (\ref{eq3.12}), (\ref{eq3.12})):
\begin{multline*}
\tilde t = t, \quad \tilde x = x + \lambda h(t_4), \quad \tilde y = y
+ \lambda g(t_4),\\
\tilde w = w +\biggl( k(t_4) - \frac{1}{2} x\dot g(t_4) - y \dot
h(t_4) \biggr)\lambda - \biggl( \dot g(t_4) h(t_4) + 2g(t_4)
\dot h(t_4) \biggr) \frac{\lambda^2}{4}.
\end{multline*}
\end{description}
The formulas for the group transformations for $N=3$ (the PKP equation
itself) are somewhat more complicated. They were given in
Ref.~\cite{ref18} and we do not reproduce them here.

The Virasoro algebra induces an arbitrary re\-par\-a\-met\-rization of
time $t_N$ (for all $N$), that is compensated for by a redefinition of
the other variables. The transformation induced by $X(h)$ is a
transition to a moving frame, moving along the $x$-axis with arbitrary
acceleration. For $h = \const$ it is a translation, for $t = t_N$ a
Galilei transformation.

\subsection{Restriction to integrable equations in 2-dimensions}
Finally, let us mention that the situation is quite different for
two-dimensional reductions of the equations of the KP hierarchy.

The Lie points symmetry group also gets reduced. The Virasoro
subalgebra, corresponding to arbitrary reparametrizations of time,
reduces to time translations and dilations only. The Kac-Moody
subalgebra in some case reduces to a finite dimensional subalgebra.
If any infinite dimensional subalgebra survives, it only corresponds
to the fact that certain functions can be added to solutions.

Let us look at individual cases.

\noindent $N=3$.

The potential KP equation (\ref{eq2.20}) has just 3 inequivalent
reductions\cite{ref17,ref18}, (by the translations $P = \partial_y$,
$P_0 = \partial_t$, or $P_1 = \partial_x$).

The reduction by $P_2$ leads to the once differential Korteweg-deVries equation
\begin{equation}
\biggl( w_t	- \frac{1}{4} w_{xxx} - \frac{3}{4} w_x^2 \biggr)_x = 0.
\label{eq4.21}
\end{equation}
The reduction of the symmetries (\ref{eq3.2}) is obtained by dropping
the $y$-derivatives and requesting that the coefficients in the vector
fields by $y$-independent. This amounts to setting
\begin{equation}
\ddot f = 0, \quad \dot g = 0, \quad \ddot h = 0, \quad k = 0
\label{eq4.22}
\end{equation}
in eq.~(\ref{eq3.2}),\dots ,~(\ref{eq3.2}), respectively. The
remaining symmetries are
\begin{equation}
\begin{gathered}
P_0 = T(1) = \partial_t, \quad D_1 = T(t) = t\partial_t + \frac{1}{3}
x\partial_x - \frac{1}{3} w \partial_w, \\
P_1 = X(1) = \partial_x, \quad B = X(t) = t \partial_x - \frac{2}{3}
x\partial_w, \quad U(\ell) = \ell(t) \partial_w.
\label{eq4.23}
\end{gathered}
\end{equation}
We see that $P_0$, $D_1$, $D$ and $B$ correspond to time and space
translations, dilations and Galilei boosts, respectively. The
infinite-dimensional subalgebra $U(\ell)$ is only present because of
the additional $x$-derivative in eq.~(\ref{eq4.17}). It corresponds
to the fact that if $w(x,t)$ is a solution, then so is $\tilde w(x,t)
= w(x,t) + \ell(t)$, where $\ell$ is an arbitrary function of $t$.

The reductions by $P_0$ leads to the potential Boussinesq equation
(with $y$ as time)
\begin{equation}
w_{xxxx} + 6w_x w_{xx} + w_{yy} = 0.
\label{eq4.24}
\end{equation}

The reduction of the symmetries (\ref{eq3.2}) in this case corresponds
to dropping $\partial_t$ and restricting the coefficients to be time
independent
\begin{equation}
f{''} = 0, \quad g' = h' = k' = \ell' = 0.
\label{eq4.25}
\end{equation}

The reduced symmetries are
\begin{equation}
D = T(t) = x\partial_x + 2y\partial_y - u\partial_u, \quad P_1 = X(1)
= \partial_x, \quad P_2 = Y(1) = \partial_y, \quad W(1) = y
\partial_w, \quad U(1) = \partial_w,
\label{eq4.26}
\end{equation}
so the inherited symmetry algebra is five-dimensional.

The reduction by $P_1$ leads to a linear equation $w_{yy} = 0$. The
inherited symmetries are generated by
\begin{equation}
P_0 = \partial_t, \quad D = t\partial_t + \frac{2}{3} y \partial_y -
\frac{1}{3} w \partial_w, \quad W(k) = k(t) y \partial_w, \quad
U(\ell) = \ell (t) \partial_w.
\label{eq4.27}
\end{equation}

We remark here that a reduced equation may have additional symmetries,
not inherited from the original equation.

For $N \ge 4$ the situation is quite similar. The most interesting
reductions are obtained using the translation $\partial_{t_N}$. The
inhereted symmetry algebras in this case are always four-dimensional, namely
\begin{equation}
D = x\partial_x + 2 \partial_y - w\partial_w, \quad P_1 = \partial_x,
\quad P_2 = \partial_y, \quad w = \partial_w.
\label{eq4.28}
\end{equation}
Those for the other reductions are easy to obtain and we shall not
spell them out here.

In this Section we chose to interpret symmetries as acting on
solutions. In view of the diagram of eq.~(\ref{eq2.58}) we could have
interpreted them equally well in terms of $\tau$-functions and vertex
operators.

\section{Conclusions}

The results of this article confirm a conjecture made earlier, namely
that the Lie point symmetries of integrable PDEs in 3-dimensions are
not only infinite-dimensional, but have a characteristic
Kac-Moody-Virasoro structure. So far we have shown that this is true
for every equation in the KP hierarchy. In doing so we have obtained
the Lie symmetry algebras of all the equations explicitly.

If the conjecture is born out, it will provide a very simple and
algorithmic ``integrability indicator'' that could rule out, or
confirm the possible existence of Lax pairs, etc., for a given
differential equation in three variables.

The converse statement cannot be made. Indeed, large families of
equations in 3 independent variables have been
constructed\cite{ref30}, that have Kac-Moody-Virasoro Lie point
symmetry algebras, but do not satisfy other integrability criteria.

Work on an extension of the results of this paper to other hierarchies
of integrable equations is in progress. In particular we note that an
integrable system with two continuous and one discrete independent
variable, namely the two-dimensional Toda lattice~\cite{ref31,ref32}
also has a Kac-Moody-Virasoro symmetry algebra~\cite{ref33} and hence
fits into the present conjecture.

\section*{Acknowledgements} We thank Professor L. A. Dickey for his
interest in this work and for helpful comments. Discussions with
Professors D. Levi and M. A. Olshanetsky were much appreciated. The
research of one of the authors (P.W.) was partially supported by
research grants from NSERC of Canada and FCAR du Qu\'ebec. A. Yu. O.
thanks the Centre de recherches math\'ematiques, Universit\'e de
Montr\'eal, where this study was performed, for its hospitality.

\end{document}